\documentstyle[twoside,fleqn,espcrc2,epsf]{article}

\newcommand{\bref}[1]{(\ref{#1})} 
\newcommand{\ct}[1]{\cite{#1}}

\newcommand{\be}{\begin{equation}} 
\newcommand{\bea}{\begin{eqnarray}} 
\newcommand{\ee}{\end{equation}} 
\newcommand{\eea}{\end{eqnarray}} 
\def\NPB#1#2#3{ {Nucl.{\,}Phys.{\,}}{\bf B{#1}} ({#3}) {#2}} 
\def\PLB#1#2#3{ {Phys.{\,}Lett.{\,}}{\bf {#1}B} ({#3}) {#2}} 
\def\PRL#1#2#3{ {Phys.{\,}Rev.{\,}Lett.{\,}}{\bf  {#1}} ({#3}) {#2}} 
\def\PRD#1#2#3{ {Phys.{\,}Rev.{\,}}{\bf D{#1}} ({#3}) {#2}}

\def\lsim{\mathrel{\rlap{\lower4pt\hbox{\hskip1pt$\sim$}}
    \raise1pt\hbox{$<$}}}
\def\gsim{\mathrel{\rlap{\lower4pt\hbox{\hskip1pt$\sim$}}
    \raise1pt\hbox{$>$}}}

\newcommand{\AmS}{{\protect\the\textfont2
  A\kern-.1667em\lower.5ex\hbox{M}\kern-.125emS}}

\title{\vspace{-3truecm}
{\small
\rightline{NIKHEF 97-036}
\rightline{CCNY-HEP-97/7}
\rightline{hep-lat/9709059}}
\vspace{1truecm}
Monte Carlo simulations of the CP$^3$ model and U($1$) gauge theory
in the presence of a $\theta$ term}

\author{Jan C.\ Plef\/ka\address{NIKHEF, P.O. Box 41882, 1009 DB Amsterdam, 
        The Netherlands} and
        Stuart Samuel\address{Department of Physics, City College of New York,
        New York, NY 10031, USA}}
       
\begin{document}

\begin{abstract}
A $\theta$ term, 
which couples to 
topological charge, 
is added to the two-dimensional 
lattice $CP^{3}$ model 
and $U(1)$ gauge theory.  
Monte Carlo simulations are performed and 
compared to strong-coupling character expansions.   
In certain instances,  
a flattening behavior occurs 
in the free-energy at sufficiently large $\theta$, 
but the effect is an artifact 
of the simulation methods.  
\end{abstract}

\maketitle

Following the discovery of instanton solutions in
four dimensional Yang-Mills theories \ct{bpst75},
the importance of adding a $\theta$ term 
$ 
  S_{\theta} = 
   g^2 \theta \int d^{4} x  F_{\mu \nu}^a 
      \tilde F^{\mu \nu}_a (x) / (32 \pi^2) 
$ 
to the action was realized \ct{cdg76,jr76}.
Since $S_{\theta}$
breaks parity, 
time-reversal invariance 
and CP symmetry  
when $\theta \ne 0$ or $\theta \ne \pi$, 
the strong interactions 
explicitly violate these symmetries 
for $0 < \theta < \pi$. The physically effective
$\theta$ angle is bounded experimentally by $\theta_{eff} 
\lsim 10^{-9}$ 
\ct{baluni79,cdvw79}. The question of how $\theta_{eff}$ 
can naturally be so
small constitutes the strong $CP$ problem in QCD.

Due to the complexity of the problem a preliminary study
of simpler systems on the lattice is useful. A class of
such systems are the two dimensional 
CP$^{N-1}$ models \ct{dlv78,witten79b}, 
which have many features in common with four dimensional 
Yang-Mills theory.

Let us start with a general analysis of simulating systems with
$\theta$ terms.
For the lattice $U(1)$ gauge theory and the $CP^{N-1}$ model, 
the local topological density $\nu_p $ is defined via  
$\nu_p \equiv \log \left( { U_p } \right) / (2\pi)$, 
where $U_p$ is the product of the $U(1)$ link phases around
the plaquette p and where 
$ -\pi < \log \left( { U_p } \right) \le \pi$.  
The total topological charge $Q$ is given by 
$Q = \sum_p \nu_p$.  
The theta term $S_{\theta \ {\rm term}}$ is $i \theta Q$,  
that is, \ct{bl81} 
\be 
  S_{\theta \ {\rm term}} = 
  {{ i\theta } \over {2\pi} } \sum_p 
   \log \left( { U_p } \right) 
\quad . 
\label{eq3p2} 
\ee 
Eq.\ \bref{eq3p2} is the lattice analog 
of the continuum $\theta$-term action 
$i { {\theta} \over {2 \pi} } \int d^2 x F_{01}$. 

Let $f( \theta )$ be the difference 
between the free energy ${\cal F} ( \theta )$ 
of a system with a $\theta$ term 
and the free energy of a system with $\theta = 0$: 
\be
  f( \theta ) = {\cal F} ( \theta ) - {\cal F} ( 0 )
\quad . 
\label{eq2p3} 
\ee 
Typically, 
$f( \theta )$ is an increasing function of $\theta$ 
for $0 \le \theta \le \pi$.  
For a fixed volume $V$, 
let $P(Q)$ be the probability of having 
a configuration with topological charge $Q$ 
in the system.  
The free energy difference $f( \theta )$ 
is then constructed from $P(Q)$ 
using   
\be
  \exp {( - V f( \theta ) )} = 
   \sum_{Q} P(Q) \exp {( i \theta Q )} 
\quad . 
\label{eq2p4} 
\ee 
Normally  
$P(-Q) = P(Q)$,  
so that $f( -\theta ) = f( \theta )$.  

In a Monte Carlo simulation, 
an approximation $f_{MC}( \theta )$ 
to $f( \theta )$ is obtained 
by using a measured $P_{MC}(Q)$ in lieu of $P(Q)$. Hence
\be
  - V f_{MC}( \theta ) ) = 
  \log { \left[ { \exp {( - V f( \theta ) )} + 
            \delta Z( \theta ) } \right] } 
\label{eq2p7} 
\ee 
where 
$
  \delta Z( \theta ) = 
   \sum_{Q} \delta P(Q) \exp {( i \theta Q )}
$.%
\footnote{The deviation between Monte Carlo measurements 
and exact results is denoted by $\delta$.}  
Since $f ( \theta ) $ is an increasing function of $\theta$, 
an accurate measurement of $f ( \theta )$ 
for $0 \le \theta < \theta_B$ 
is obtained
if 
\be 
  | \delta Z( \theta ) | \ll  \exp {( - V f( \theta_B ) )} 
\quad .       
\label{goodcrit} 
\ee 
In particular, 
since $f(0) = 0$ and 
$| \delta Z( \theta ) | \ll 1$, 
there is always a region near $\theta = 0$  
for which $f( \theta) $ can be measured 
in a Monte Carlo simulation.  
However, away from $\theta = 0$, 
Eq.\ \bref{goodcrit} implies that
for sufficiently large $V$, 
a limiting value of $\theta_B$ exists 
beyond which 
it is impossible to reliable compute $f (\theta )$.  
The value of $\theta_B$ 
depends on the statistical accuracy of the simulation.  
As $V$ gets larger, 
$\theta_B$ decreases  
unless enormous numbers of measurements  
are undertaken to reduce statistical errors.  
For large $V$, 
obtaining enough measurements becomes, 
in any practical sense, 
impossible.   
Clearly, 
it is more difficult to measure $f( \theta )$ 
throughout the entire fundamental region of $\theta$, 
as $V$ gets larger.  

\begin{figure}[t]
\epsfxsize=3in
\epsfbox{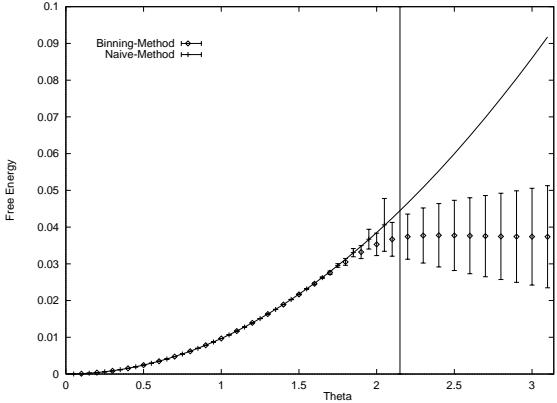}
\caption{$U(1)$ Free Energy Versus $\theta$ at $\beta=1.0$ 
for the Naive and Binning Methods.}
\end{figure}

It turns out \ct{us} that in most Monte Carlo simulations, 
there is a tendency for 
\be
  | \delta P(0) | > | \delta P(1) | > | \delta P(2) | > \dots 
\quad .    
\label{pineq} 
\ee 
Now if $| \delta P(0) |$ is much larger than the other 
$|\delta P(Q)|$
then, from Eq.\ \bref{goodcrit}, 
one deduces an estimate for $\theta_B$
\be
   f ( \theta_B ) \approx { {1} \over {V} } | \log | \delta P(0) | | 
\quad .    
\label{estthetab} 
\ee 
Since Monte Carlo results are reliable for $\theta < \theta_B$,  
\be
   f_{MC} ( \theta ) \approx  f ( \theta ) 
\quad {\rm for} \ \theta < \theta_B
\quad . 
\label{eq2p11} 
\ee 
If, in addition, $\delta P(0) > 0$, 
then  
one finds%
\be
   f_{MC} ( \theta ) \approx 
   - { {1} \over {V} }  \log \delta P(0)   
\quad {\rm for} \ \theta > \theta_B
\quad ,  
\label{eq2p12} 
\ee 
so that a constant ``flat'' behavior 
in $f_{MC} ( \theta )$ will be observed, a pure artifact of the
simulation. 
If, on the other hand, $\delta P(0) < 0$, 
then the measured $f_{MC} (\theta )$ will blow up 
for $\theta > \theta_B$. In our simulations we have observed
both types of behaviours.

\begin{figure}[t]
\epsfxsize=3in
\epsfbox{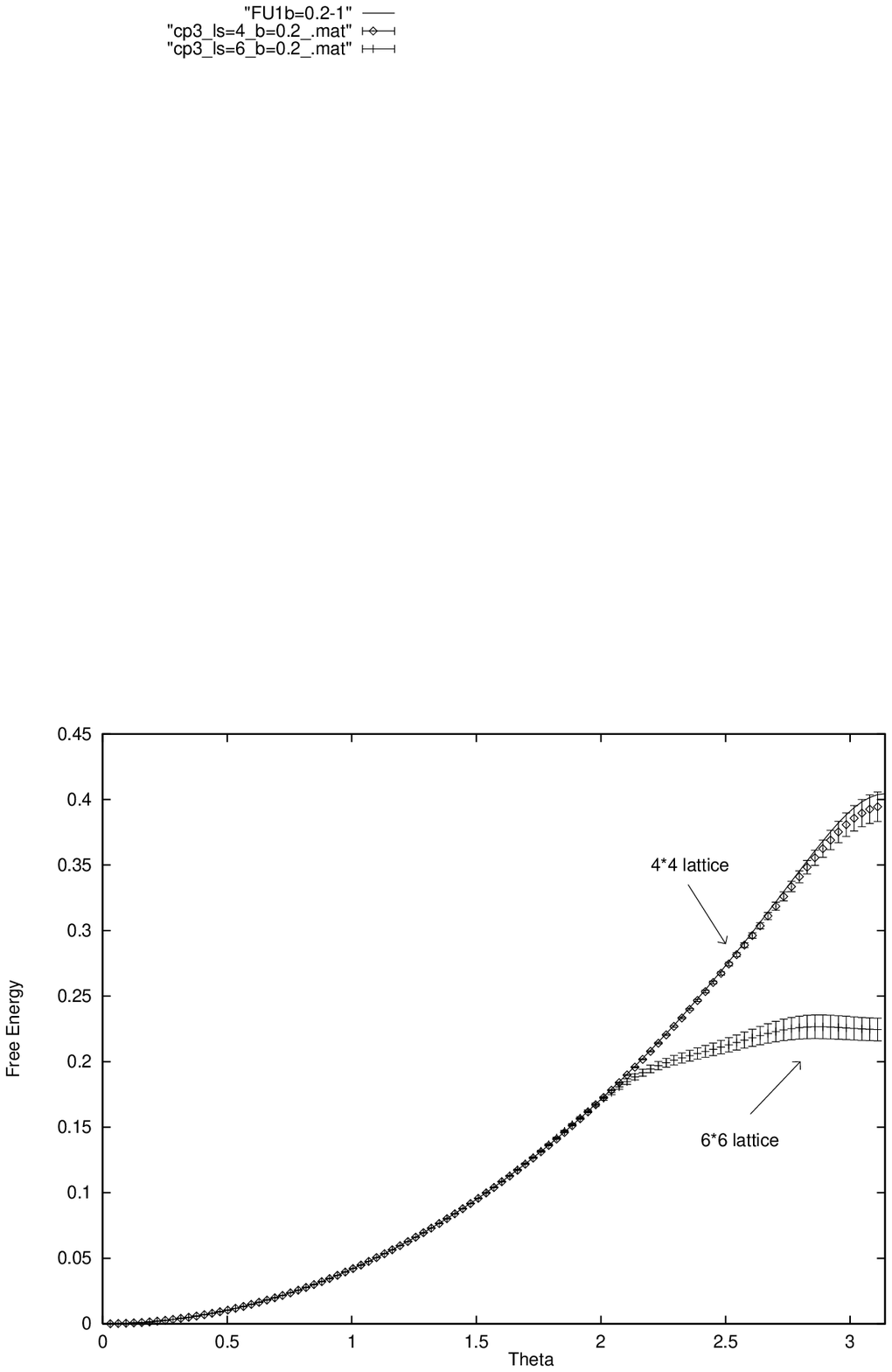}
\caption{$CP^3$ Free Energy Versus $\theta$ at $\beta=0.2$ 
on $4\times 4$ and $6\times 6$ lattices.}
\end{figure}

{}From the above discussions we see that as long as finite-size 
effects are under control, that is $\xi < V^{(1/d)}$,%
\footnote{Here,
$\xi$ is the correlation length
and $d$ is the number of dimensions of the system.}
small-volume results for the measurement of $f (\theta )$ 
are more reliable than large-volume results.
If a flattening behavior of the free energy $f( \theta)$ 
for large $\theta$ is observed, 
one should be cautious that the result is spurious.  
In particular, 
one should try to see whether $| \delta P(0) |$ 
is bigger than the other $| \delta P(Q) |$.  
Therefore the guideline emerges that if a large-volume 
simulation shows a flattening effect 
for $f( \theta)$ for $\theta$ sufficiently large, 
but a smaller-volume simulation does not, 
one should trust the smaller-volume result.  

We note that in the work of \ct{schierholz94} a flat behaviour
of the free energy was observed and attributed to a
phase transition. The results of this work \ct{us}
suggest that the flattening is a simulation effect. 

The 2-D lattice $U(1)$ gauge theory serves as an ideal testing
ground, as computer simulations can be compared to exact
analytic results \ct{wiese89}.  
Figure 1 plots the free energy versus $\theta$ 
for $\beta = 1.0$ on
a periodic $16 \times 16$ lattice for two different runs. 
The solid line is the exact
analytic result. 
Both runs have comparable statistics and agree with
the analytic results for $\theta$ less than $2.1$,
the value of the ``barrier $\theta$'' $\theta_B$. Using the 
known error $\delta P(0)$ in Eq.\bref{estthetab}
to estimate $\theta_B$, one finds $\theta_B \approx 2.05$, confirming
the above data analysis. 
The run exhibiting the anomalous flat behaviour in the free
energy for $\theta> 2.1$ in fact has a positive
$\delta P(0)$, as predicted by Eq.\ \bref{eq2p12}.

\begin{figure}[t]
\epsfxsize=3in
\epsfbox{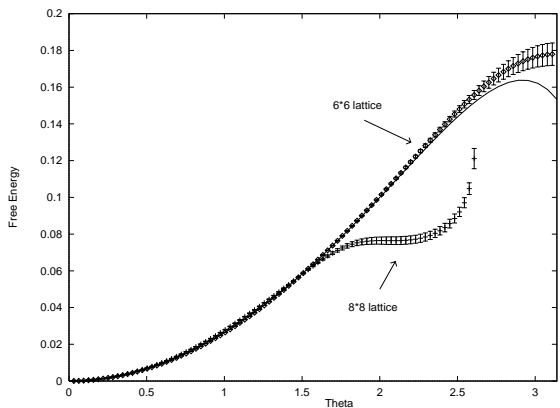}
\caption{$CP^3$ Free Energy Versus $\theta$ at $\beta=0.6$ 
on $6\times 6$ and $8\times 8$ Lattices.}
\end{figure}

For the simulations of the lattice $CP^3$ model%
\footnote{For simulations without a $\theta$ term see refs.\ in
\ct{us}}
we have employed
the ``auxiliary U(1) field'' formulation \ct{samuel83}.
Figures 2,3 and 4 show the free energy for $\beta= 0.2, 0.6$ and 
$0.7$ on $4^2,6^2$ and $8^2$ lattices. The solid line represents
the tenth-order strong-coupling character expansion 
of ref.\ \ct{ps96a}. Figures 2 and 3 show 
that simulations on smaller lattices are more reliable,
as the simulations on the larger lattices exhibit
anomalous flattening. Again the estimated $\theta_B$ for
these simulations was in good agreement with the observed
one. In the intermediate coupling
regime of $\beta=0.7$ in figure 4 the Monte Carlo data is most
likely to be trusted over the strong-coupling expansion. Curiously
for higher values of $\beta$ the MC simulations were nicely
fitted by a cosine \ct{us}, which also arises from a
 topological gas picture \ct{cdg76}.

\begin{figure}[t]
\epsfxsize=3in
\epsfbox{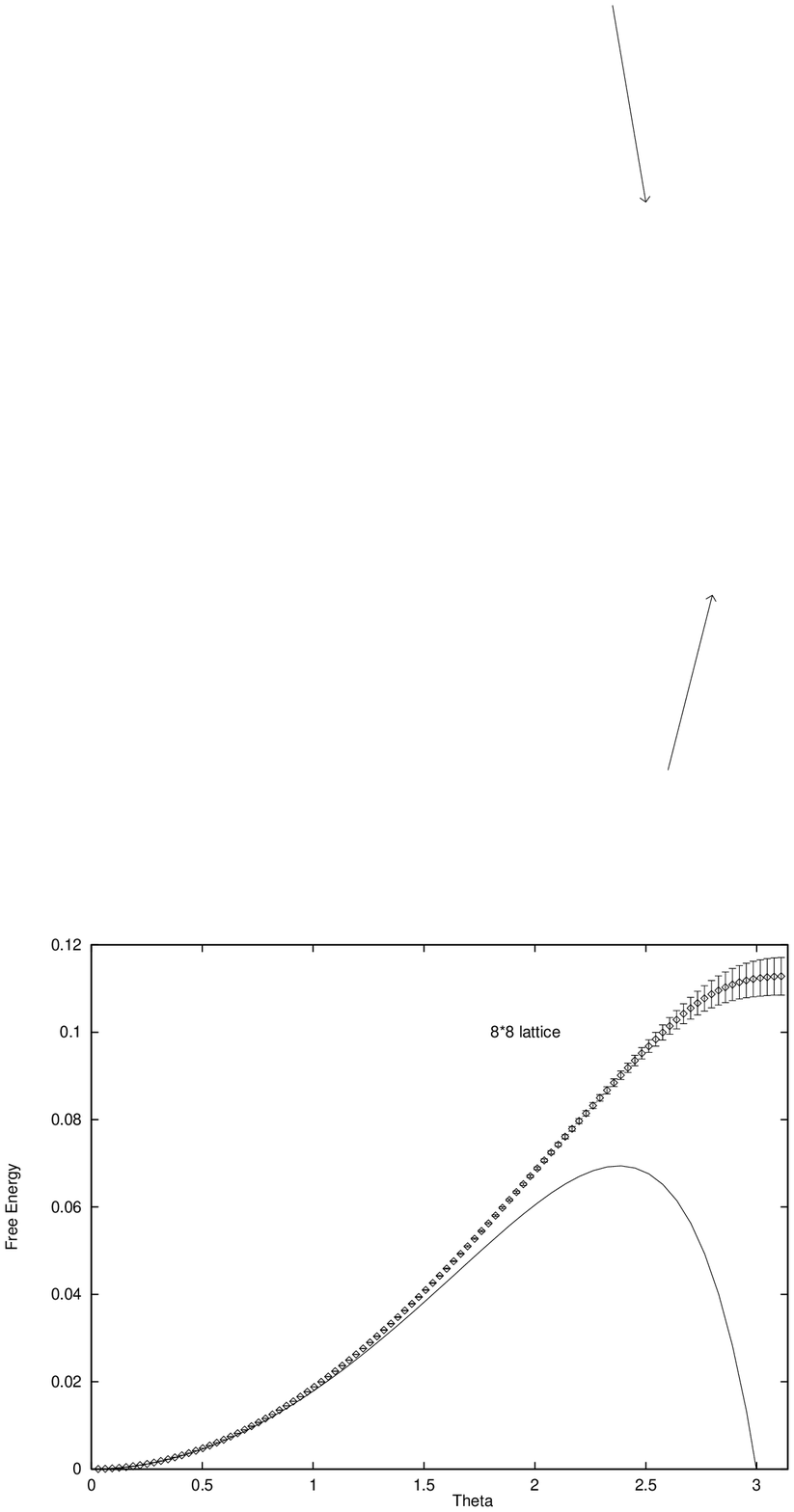}
\caption{$CP^3$ Free Energy Versus $\theta$ at $\beta=0.7$ 
on a $8\times 8$ Lattice.}
\end{figure}


\begin{thebibliography}{9}   
\bibitem{bpst75} 
A.\,Belavin, A.\,Polyakov, A.\,Schwartz and Y.\,Tyupkin, 
\PLB{59}{85}{1975}. 

\bibitem{cdg76} 
C.\,Callan, R.\,Dashen and D.\,Gross, 
\PLB{63}{334}{1976}. 

\bibitem{jr76} 
R.\,Jackiw and C.\,Rebbi, 
\PRL{37}{172}{1976}
 
\bibitem{baluni79} 
V.\,Baluni, 
\PRD{19}{2227}{1979}. 

\bibitem{cdvw79} 
R.\,Crewther, P.\,Di Vecchia, 
G.\,Veneziano and E.\,Witten, 
\PLB{88}{123}{1979}. 

\bibitem{dlv78} 
A.\,D'Adda, M.\,L\"uscher and P.\,Di Vecchia, 
\NPB{146}{63}{1978}. 

\bibitem{witten79b} 
E.\,Witten, 
\NPB{149}{285}{1979}.

\bibitem{bl81}  
B.\,Berg and M.\,L\"uscher, 
\NPB{190}{412}{1981}. 

\bibitem{schierholz94}  
S.\,Olejnik and G.\,Schierholz, 
Nucl.\,Phys.\,B (Proc.\,Suppl.) { \bf 34} (1994) 709; \\ 
G.\,Schierholz,
Nucl.\,Phys.\,B (Proc.\,Suppl.) { \bf 37A} (1994) 203.  

\bibitem{us} 
J.\,Plefka and S.\,Samuel, 
\PRD{56}{44}{1997}.

\bibitem{wiese89} 
U.-J.\,Wiese, 
\NPB{318}{153}{1989}. 

\bibitem{ps96a} 
J.\,Plefka and S.\,Samuel, 
\PRD{55}{3966}{1997}.


\bibitem{samuel83} 
S.\,Samuel, 
\PRD{28}{2682}{1983}. 
\end{thebibliography}
\end{document}